\documentstyle[12pt]{article}
\def\1{\mbox{1\hspace{-.25em}I}}

\def\R{{\rm I\hspace{-.15em}R}}

\def\b{\begin{equation}}
\def\e{\end{equation}}
\def\bee{\begin{enumerate}}
\def\eee{\end{enumerate}}

\title{N=1 super-symmetry Lagrangian in de Sitter space}

\author{ M. R. Masoumi Nia \thanks{e-mail: m.masouminia@ut.ac.ir} }

\begin{document}
\maketitle 
\begin{abstract}

Previously, Pahlavan, Rouhani and Takook have introduced a novel $N=1$ super-symmetric algebra in de Sitter space-time. This paper is an attempt to build a proper $N=1$ super-symmetric field theory of classical level in the de Sitter space. The generators, gauge transformations and different fields in a 5-dimensional ambient space notation are defined and corresponding super-space and super-fields are introduced. Finally, the $N=1$ super-symmetry Lagrangian in the de Sitter ambient space notation has been derived.

\end{abstract}

\section{Introduction}

Searching for a fundamental theory to describe the nature is the sole purpose of all theoretical physical attempts. This quest brought us from the simplest models to currently most complicated theories in an indirect yet unchanged path, i.e. extension of the symmetries. This extension led us to the Standard Model, then to the Grand Unification theory and afterwards to the idea of the super-symmetry. This may be not the end of the route but it is the best we can produce by the concept of point-particle in Minkowski space-time. Super-symmetry has been surveyed from many points in the Minkowski space-time \cite{il08,bi01,ma08}. However, the extension of the symmetries of the Minkowskian space-time (Poincare group $T^4\times SO(1,3))$ lead us to the de Sitter group $SO(1,4)$, anti-de Sitter group $SO(2,3)$ and conformal group $SO(2,4)$.

Today, considering latest discoveries of the modern astrophysical data, we believe that our world in first approximation is in a de Sitter phase. Therefore, it is important to find a formulation of de Sitter quantum field theory with the same level of completeness and rigour as its Minkowskian counterpart. This formalism was first developed by Bros-Gazeau-Moschella for scalar field $(s=0)$ and generalized to other fields $(s=\frac{1}{2}, 1, \frac{3}{2}, 2)$  by Gazeau et al., based on a theoretical rigorous group approach combined with a suitable adaptation of the Wightman-Garding axiomatic and the analyticity of the two-point function in the ambient space notation. It has been shown that the massive and massless conformally coupled scalar fields $(s=0)$ in the de Sitter space correspond to the principal and complementary series representations of the de Sitter group, respectively \cite{brgamo94,brmo96}. The massive fields with a variety of spins $(s=\frac{1}{2}, 1, \frac{3}{2}, 2)$ are associated with the principal series representation of de Sitter group \cite{ta97,ta96,bagamota01,gata00,gagata03,gasp97,taazba12}, whereas massless fields correspond to the lowest of the discrete
series representation of the de Sitter group \cite{ta97,gareta00,gagarota08,taaz12,ta09,ta99,derotata08,anla98}.

Historically, some questions are usually put forth for the non-existence of super-symmetry models with a positive cosmological constant, i.e. super-symmetry in the de Sitter space. Such arguments are often based on the non-existence of Majorana spinors for $SO(1,4)$ \cite{pini85,dehe00}. Pilch et al. have shown that for every spinor and its independent charge-conjugate, de Sitter super-gravity can be established with even $N$ \cite{pini85}.

In their previous work, \cite{PRT05}, Pahlavan et al. have introduced a new definition of complex conjugate spinor in the de Sitter ambient apace notation \cite{ta97,bagamota01,morota05}, showing that a super-symmetric algebra can be establish in de Sitter space for any odd or even values of $N$ Now, it seems reasonable to aim for the construction of a super-symmetric field theory with $N=1$ in the de Sitter space-time.

In this paper, after a brief introduction of the utilized notation, the aforementioned $N=1$ super-algebra in the de Sitter space has been reproduced. For some technical reasons, related to the chirality of the intended model (see $sec. 4$), a transformation to the corresponding Weyl representation is made and then, the proper super-space and super-fields in the de Sitter space has been constructed. Afterwards, defining relevant kinetic terms, $N=1$ super-symmetric de Sitter Lagrangian is obtained.

\setcounter{equation}{0}
\section{Notation}

The de Sitter space-time is a solution of the vacuum Einstein equation with positive cosmological constant. It is conveniently
seen as a hyperboloid embedded in a five-dimensional Minkowski space \cite{brgamo94}
	 \b
	 dS_4=\{x \in \R^5 ;x^2=\eta_{\alpha\beta} x^\alpha x^\beta
	 =-H^{-2}\},\;\;
     \alpha,\beta=0,1,2,3,4 , \label{eq2.1}
     \e
where $\eta_{\alpha\beta}=\mbox{diag}( 1,-1,-1,-1,-1 )$. It has constant curvature $12H^{2}$ and reproduces the Minkowski space-time in the zero curvature limit $H=0$ (i.e. when the radius $H^{-1}$ tends to infinity).
The de Sitter metrics reads
	 \b
	 dS^2=\eta_{\alpha\beta}dx^{\alpha}dx^{\beta}\mid_{x^2=-H^{-2}}=g_{\mu\nu}^{dS}dX^{\mu}dX^{\nu};\;\;
	 \mu,\nu=0,1,2,3, \label{eq2.2}
	 \e
where $X^\mu$'s are the $4$ space-time intrinsic coordinates on the $dS_4$ hyper-surface.

A 10-parameter group $SO(1,4)$ is the kinematical group of the de Sitter universe. In the limit of $H=0$, de Sitter group reduces to the Poincare group. One introduces the Invariant operators for the $SO(1,4)$ group as follows:
     \b
     Q_1=-\frac{1}{2}L_{\alpha\beta} L^{\alpha\beta} ,\;\;\;
	 Q_2=-W_{\alpha}W^{\alpha};\;\;\;W_{\alpha}=-\frac{1}{8}
	 \epsilon_{\alpha\beta\gamma\delta\eta}L^{\beta\gamma} L^{\delta\eta}, \label{eq2.3}
	 \e
where $L_{\alpha\beta}$ are the infinitesimal generators of the de Sitter space-time \cite{PRT05}:
	 \b
	 \lbrack L_{\alpha\beta}, L_{\gamma\delta}\rbrack
	 = -i(\eta_{\alpha\gamma}L_{\beta\delta}+\eta_{\beta\delta}
	 L_{\alpha\gamma}-\eta_{\alpha\delta}L_{\beta\gamma}-\eta_{\beta\gamma}
	 L_{\alpha\delta}). \label{eq2.4}
	 \e
One may represent $L_{\alpha\beta}$ as
	 \b
	 L_{\alpha\beta}= S_{\alpha\beta} + M_{\alpha\beta} , \label{eq2.5}
	 \e
with $S_{\alpha\beta}$ and $M_{\alpha\beta}$ being Spin and orbital contributions respectively:
 	\b
 	S_{\alpha\beta}=-{i\over 4}\lbrack\gamma_{\alpha},\gamma_{\beta}\rbrack , \label{eq2.6}
 	\e
	\b
	M_{\alpha\beta}  
	= -i(x_{\alpha}\partial_{\beta}-x_{\beta}\partial_{\alpha})
	= -i(x_{\alpha}\partial_{\beta}^{T}-x_{\beta}\partial_{\alpha}^{T}). \label{eq2.8}
	\e
$\gamma_{\alpha}$ are a set of $4\times4$ matrices which can be identified as the generators of Clifford algebra;
	 $$
	 \gamma^0=\left( \begin{array}{clcr} \1 & \;\;0 \\ 0 &-\1 \\ \end{array} \right)
	 ,\;\;\; \gamma^4=\left( \begin{array}{clcr} 0 & \1 \\ -\1 &0 \\ \end{array} \right),
	 $$
	 \b
	 \gamma^1=
	 \left( \begin{array}{clcr} 0 & i\sigma^1 \\ i\sigma^1 &0 \\    \end{array} \right)
     ,\;\;\; \gamma^2=\left( \begin{array}{clcr} 0 & -i\sigma^2 \\ -i\sigma^2 &0 \\  \end{array} \right)
     ,\;\;\;  \gamma^3=\left( \begin{array}{clcr} 0 & i\sigma^3 \\ i\sigma^3 &0 \\   \end{array} \right). \label{eq2.9}
     \e
Additionally, in writing $(2.6)$ and $(2.7)$, the following definitions have been utilized:
	 \b
	 \partial_{\alpha} = {\partial \over \partial x^{\alpha}} \; , \;\;\;
	 \partial_{\alpha}^T=\Theta^\beta_\alpha \partial_\beta=\partial_{\alpha}+H^2x_{\alpha}x\cdot\partial , \label{eq2.10}
	 \e
	 \b
	 \lbrace\gamma^{\alpha},\gamma^{\beta}\rbrace
	 =2\eta_{\alpha\beta} \1 , \;\;\; (\gamma^{\alpha})^\dagger 				
	 =\gamma^0 \gamma^\alpha \gamma^0 , \;\;\; (\gamma^{\alpha})^T
	 =\gamma^4 \gamma^2 \gamma^\alpha \gamma^2 \gamma^4 , \label{eq2.11}
	 \e
where $\Theta_{\alpha\beta}=\eta_{\alpha\beta}+H^2x_{\alpha}x_{ \beta}$ represents a transverse projection tensor ($\Theta_{\alpha\beta}\; x^{\alpha}=\Theta_{\alpha\beta} \; x^{\beta}=0$).
 
One may introduce the transformation $\Lambda \in SO_0(1,4)$ for the elements of $Sp(2,2)$ group through 
\cite{ta97,bagamota01}
	 \b
	 x^\alpha \longrightarrow x'^\alpha
	 = \Lambda^{\alpha}_{\beta}x^{\beta} \; ; \;\;\; \Lambda\in SO(4,1) ,
	 \e
 	 \b
	 \Lambda^{\alpha}_{\beta}(g)=\frac{1}{4}tr(\gamma^{\alpha}
	 g\gamma_{\beta}g^{-1}),\;\;\;
	 \Lambda^{\alpha}_{\beta}\gamma^{\beta}=g\gamma^{\alpha}g^{-1} ,
	 \e
	 \b
 	 \psi(x) \longrightarrow \psi '(x') = g\psi(x)\; ; \;\;\; g\in Sp(2,2).
 	 \e
Therefore, using ${\overline \psi}(x)\equiv \psi^{\dag}(x){\gamma^0}{\gamma^4}$ and $g^{-1} = \gamma^0 g^\dagger \gamma^0 $, the transformation law for $\overline{\psi}(x)$ reads
	 \b
	 {\overline\psi}(x)  \ \longrightarrow\ {\overline \psi}^{\prime}(x')
	 ={\overline \psi}\bigr(x\bigr)[-\gamma^4 g^{-1} \gamma^4] .
	 \e
Additionally, to further our quest for construction of a viable $N=1$ super-algebra in de Sitter space-time and consequently 
its respective field theory, one needs to redefine the concept of charge conjugation spinor, $\psi^c$, in the ambient space notation, as it has been done in \cite{morota05}, also look \cite{PRT05} :
	 \b
	 \psi^{c}=\eta_c C(\gamma^4)^T(\bar \psi)^T  , 
	 \e
with $\eta_c$ as an arbitrary phase factor. The commutation relations between the charge conjugation operator, $C$, and
$\gamma_{\alpha}$ matrices are as follows:
	 $$
	 C\gamma^{0}C^{-1}=\gamma^0 \; , \;\;\; C\gamma^{1}C^{-1}=\gamma^1 \; , \;\;\;
	 C\gamma^{2}C^{-1}=-\gamma^2 \; ,$$ \b \;\;\; C\gamma^{3}C^{-1}=\gamma^3 \; , \;\;\;
	 C\gamma^{4}C^{-1}=-\gamma^4 .
	 \e
In \cite{morota05}, it has been established that the simplest choice for the charge conjugation operator is $C=\gamma^2\gamma^4$, which ultimately satisfies its characteristic equation
	 \b
	 C=-C^{-1}=-C^T=-C^\dag .
	 \e
Comparing to $\psi$, the adjoint spinor ${\overline \psi}(x)$, transforms in a different manner, under de Sitter transformation. Yet, it can be shown easily that $\psi^c$ transforms in the same way as $\psi$:
	 \b
	 \psi^c(x) \longrightarrow \psi'^c(x')=g(\Lambda)\psi^c(x).
	 \e
$\psi$ and $\psi^c$ have the same charge but with the wrong sign;
	 \b
	 \psi^c = C \gamma^0 \psi^* \; ; \;\;\; ({\psi}^c)^c
	 = C\gamma^0(C \gamma^0 \psi^*)^*=\psi,
	 \e
therefore, If $\psi$ is considered to describe the presence of a dS-Dirac "particle" with charge $q$, $\psi^c$ is a sound candidate to represent its dS-Dirac "antiparticle" counterpart.

\setcounter{equation}{0}
\section{N=1 Super-algebra}

The first step in applying a super-symmetric extension on the de Sitter group would be the introduction of a complete set of spin-generating operators, $Q^n_i$ with spinor index $i=1,2,3,4$ and internal symmetry index $n=1,...,N$. In \cite{PRT05}, this rather complex task has been carried out, in an impressive manner. 

A de Sitter super-algebra has been created, using the following generators:
	\begin{itemize}
    	\item  $L_{\alpha\beta}$, the generator of infinitesimal transformation of the de Sitter group.
    	\item  $A_{mn}$, the generators of the internal symmetry which are defined as
		$$A_{mn}=-A_{nm}\;;\;\;\; m,n=1,2....N.$$
		\item  A four-component dS-Dirac spinor generator, $$Q^n_i,\;\;i=1,2,3,4,\;\;n=1,...,N.$$
	 \end{itemize}
Thus, the desired  super-algebra forms as follows:
	$$ [L_{\alpha\beta}, L_{\gamma\delta}] =
	-i(\eta_{\alpha\gamma}L_{\beta\delta}+\eta_{\beta\delta}
	L_{\alpha\gamma}-\eta_{\alpha\delta}L_{\beta\gamma}-\eta_{\beta\gamma}
	L_{\alpha\delta}),$$
	$$ [A_{mn},A_{rs}]=
	-										   																				 i(\varepsilon_{mr}A_{ns}+\varepsilon_{ns}A_{mr}-\varepsilon_{ms}A_{nr}-\varepsilon_{nr}A_{ms}),
	$$
	$$
	[L_{\alpha\beta}, A_{mn}] =0 \; , \;\;\; 																				 {[Q^{m}_{i},A^{rs}]}={-(\varepsilon^{rl}Q^{r}_{i}-\varepsilon^{rp}Q^{s}_{i})} ,
	$$
	$$
	[Q^{m}_{i},L_{\alpha\beta}]=(S_{\alpha\beta}Q^{m})_{i} \; ,\;\;\;
	[\tilde Q^{m}_{i},L_{\alpha\beta}]=-(\tilde Q^{m} S_{\alpha\beta})_{i} ,
	$$
	$$
	\{Q^{m}_{i},Q^{n}_{j}\}=\varepsilon^{mn}\left(S^{\alpha\beta}\gamma^4
	\gamma^2\right)_{ij}L_{\alpha\beta}+ \left(\gamma^4
	\gamma^2\right)_{ij}A^{mn} .
	$$
The matrix elements of $\varepsilon$ determine the structure of the internal symmetry of the group. With proper adjustments it must reproduce de Sitter super-symmetry algebra with even $N$, already studied by Pilch et al. \cite{pini85}. Consequently, $\varepsilon$ may be obtained explicitly. On the other hand, this super-algebra is also closed for odd $N$, due to the definitions of complex conjugation spinor and charge conjugation in ambient space notation \cite{ta97,bagamota01,morota05}.
	 \b
	 \tilde{Q}_i=\left(Q^T \gamma^4 C\right)_i=\bar{Q^c_i},
	 \e
where, it can be shown that $\tilde{Q}\gamma^4 Q$ is a scalar field under the de Sitter transformation. This concludes a new
form of de Sitter super-algebra in the ambient space notation for $N = 1$, with $A_{11} = 0,\varepsilon = 1$:
	 \b [L_{\alpha\beta}, L_{\gamma\delta}] =
	 -i(\eta_{\alpha\gamma}L_{\beta\delta}+\eta_{\beta\delta}
	 L_{\alpha\gamma}-\eta_{\alpha\delta}L_{\beta\gamma}-\eta_{\beta\gamma}
	 L_{\alpha\delta}),\e
	 \b [Q_i,L_{\alpha\beta}]=(S_{\alpha\beta}Q)_i \; , \;\;\; [\tilde Q_i,L_{\alpha\beta}]=
	 -(\tilde Q S_{\alpha\beta})_i , \e	
	 \b \{Q_i,Q_j\}=\left(S^{\alpha\beta}\gamma^4
	 \gamma^2\right)_{ij}L_{\alpha\beta}.\e
This can be proved, using the generalized Jacobi identities.

\setcounter{equation}{0}
\section{N=1 super-space and super-fields in de Sitter space}

To construct a super-symmetric model, it is instructive to define a proper $N=1$ super-space on the de Sitter space-time. This super-space contains $4$ grassmannian coordinates $(\theta,\tilde{\theta})$\cite{il08,bi01} and de Sitter space-time ambient space coordinates $x^\alpha$. Meanwhile, it is often convenient to use spinors with definite chirality. Therefore, it is better to work out $N=1$ super-algebra on Weyl representation. By defining
	 \b
	 \sigma^\alpha =(1 , i\sigma^1 , -i\sigma^2 , i\sigma^3 , -1),
	 \e
and
	 \b
	 \sigma^{\alpha\beta} = -{{i} \over 4} (\sigma^{[\alpha }\bar{ \sigma}^{\beta]})^i_j ,
	 \e
it can be shown that the $N=1$ super-algebra in Weyl spinors representation can be written as follows (see the appendix A of the reference \cite{PRT05}):
	 \b
	 [L_{\alpha\beta}, L_{\gamma\delta}] =
	 -i(\eta_{\alpha\gamma}L_{\beta\delta}+\eta_{\beta\delta}
	 L_{\alpha\gamma}-\eta_{\alpha\delta}L_{\beta\gamma}-\eta_{\beta\gamma}
	 L_{\alpha\delta}),
	 \e
 \b
	 \{Q_i,\tilde{Q}_j\}_{Weyl}=\sigma^{\alpha\beta}_{ij}L_{\alpha\beta},
	 \e
 \b
	 \{Q_i,Q_j\}_{Weyl}=0 \; , \;\;\;
	 \{\tilde{Q}_i,\tilde{Q}_j\}_{Weyl}=0 ,
	 \e
	 \b
	 [Q_i,L_{\alpha\beta}]_{Weyl}=+ \sigma_{\alpha\beta}^{ij}Q_j \; , \;\;\;
	 [\tilde{Q}_i,L_{\alpha\beta}]_{Weyl}=-\sigma_{\alpha\beta}^{ij}\tilde{Q}_j ,
	 \e	 	
where $Q_i$ and $\tilde{Q}_i $, $i=1,2$ are super-symmetric spinorial generators in Weyl representation. Note that in dS-Dirac representation, a 4-component spinorial generator (corresponding to four grassmannian coordinates) has been used. Here, in Weyl representation, we are using two 2-component spinorial generators ($Q_i$ as generators of transformations on grassmannian coordinates $\theta_i, i=1,2$ and $\tilde{Q}_j$ as generators of transformations on grassmannian coordinates $\tilde{\theta}_j, j=1,2$) which are related through complex conjugation. From here on, we will always use notations $Q$ and $\tilde{Q}$ for spinorial generators in Weyl representation.

Now, one can introduce $(x^{\alpha},\theta,\tilde{\theta})$ as an element of $N=1$ super-space on the de Sitter space-time. A finite group element of super-symmetric transformation on the de Sitter space may be defined as fallows \cite{il08}:
	 \b
	 G(x^{\alpha},\theta,\tilde{\theta})
	 =exp \{i\theta Q +i\tilde{\theta}\tilde{Q} \} exp \{ -ix^{\alpha}\sigma^{\beta}L_{\alpha\beta}\} ,
	 \e
where
	 $$
	 \theta Q = \theta^i Q_i .
	 $$
One other choice for defining the group element is:
	 \b
	 G'(x^{\alpha},\theta,\tilde{\theta}) =
	 exp \{ i\theta Q \} exp \{ i \tilde{\theta}\tilde{Q} \} exp \{-ix^{\alpha}\sigma^{\beta}L_{\alpha\beta}\}   .
	 \e
The representations $(4.7)$ and $(4.8)$ are equivalent (but not the same):
	 \b
	 G(x^{\alpha},\theta,\tilde{\theta}) =
	 G'(x^{\alpha}+ i\theta \sigma^{\alpha} \tilde{\theta} + ... ,\theta,\tilde{\theta}) .
	 \e

A function $F(x^{\alpha},\theta ,\tilde{\theta})$ is defined on the super-space. Considering the properties of the Grassmann algebra, each super-field is a polynomial \cite{il08}.  So, the most general form of a super-field in $(4.7)$ is
	 \b
	 F(x^{\beta} ,\theta,\tilde{\theta})=A(x^{\beta}) + \theta\psi(x^{\beta})
	 + \tilde{\theta}\tilde{\chi}(x^{\beta})
	 + \theta\theta m(x^{\beta}) + \tilde{\theta}\tilde{\theta}n(x^{\beta})
	 + \theta\sigma^{\alpha}\tilde{\theta}\nu_{\alpha}(x^{\beta})
	 + \theta\theta\tilde{\theta}\tilde{\lambda}(x^{\beta})
	 $$$$
	 + \tilde{\theta}\tilde{\theta}\theta\rho(x^{\beta})
	 + \theta\theta\tilde{\theta}\tilde{\theta}D(x^{\beta}) ,
	 \e
where $\theta\theta = {1 \over 2}(\theta^1 \theta^2 - \theta^2 \theta^1)$. The coefficient functions $A,m,n$ and $D$ are scalar fields, $\psi,\chi,\lambda$ and $\rho$ are spinor fields and $\nu_{\alpha}$ is a vector boson field. A simple explicit form of the super-symmetric generators, which satisfy super-algebra $(4.3-6)$, on super-field $(4.10)$ can be defined as :
	 \b
	 Q_i F \equiv \left(
	 - i \sigma^{\alpha}_{ij}x_{\alpha} {\partial \over \partial \theta_j}
	 + \sigma_{ij}^{\alpha} \tilde{\theta}_j \partial_{\alpha}^T
	 \right) F ,
	 \e
	 \b
	 \tilde{Q}_i F \equiv \left(
	  i \sigma^{\alpha}_{ij} x_{\alpha} {\partial \over \partial \tilde{\theta}_j}
	 - \theta_j \sigma_{ji}^{\alpha}  \partial_{\alpha}^T
	 \right) F ,
	 \e
	 \b
	 L_{\alpha\beta} =
	 - i(x_{\alpha} \partial_{\beta}^T - x_{\beta} \partial_{\alpha}^T)
	 + i \sigma_{\alpha\beta}^{ij} (\theta_i {\partial \over \partial \theta_j}
	 + \tilde{\theta}_i {\partial \over \partial \tilde{\theta}_j}) .
	 \e
A general super-symmetric transformation will be written in the following form:
	 \b
	 \delta_{\epsilon , \tilde{\epsilon} , y^{\alpha}} F(x^{\beta} ,\theta,\tilde{\theta})
	 = ( i \epsilon Q + i \tilde{\epsilon} \tilde{Q} - i y^{\alpha} \sigma^{\beta} L_{\alpha\beta} )
	 F(x^{\beta} ,\theta,\tilde{\theta}),
	 \e
where $\epsilon$, $\tilde{\epsilon}$ and $y^{\alpha}$ are infinitesimal parameters, i.e. $\epsilon^i \epsilon^j = \tilde{\epsilon}^i \tilde{\epsilon}^j = y^{\alpha} y^{\beta} = 0$. So, through definition
	 \b
	 \delta_{\epsilon , \tilde{\epsilon} , y^{\alpha}} [D_i F(x^{\beta} ,\theta,\tilde{\theta})] =
	 D_i [ \delta_{\epsilon , \tilde{\epsilon} , y^{\alpha}} F(x^{\beta} ,\theta,\tilde{\theta})]
	 \e
(idem for $\tilde{D}_i$), we can define covariant derivatives as
\b
	 D_i =
	 \sigma^{\alpha}_{ij} x_{\alpha} {\partial \over \partial \theta_j}
	 + i \sigma_{ij}^{\alpha} \tilde{\theta}_j \partial_{\alpha}^T ,
	 \e
	 \b
	 \tilde{D}_i =
	 \sigma^{\alpha}_{ij} x_{\alpha} {\partial \over \partial \tilde{\theta}_j}
	 + i\theta_j \sigma_{ji}^{\alpha}  \partial_{\alpha}^T .
	 \e
	
In order to extract the possible physical consequences of super-symmetry, we must construct the representations of the super-algebra in terms of one-particle states, i.e. one-particle "supermultiplets". For $N=1$ super-algebra in the de Sitter space, $(4.3-6)$, excluding gravitation, two type of supermultiplets are required \cite{il08,bi01}:
	\begin{itemize}
    	\item  "Chiral supermultiplets", which are contain one spin $1 \over 2$ Weyl fermion field and one scalar field.
    	\item  "Vector supermultiplets", which are contain one spin $1 \over 2$ Weyl fermion field and one spin $1$ vector field.
	 \end{itemize}

In case of building a super-symmetric $N=1$ model, it is necessary to represent supermultiplets in term of super-fields, which are defined on the irreducible representations of the super-symmetry group.  Fortunately, such a this operation is possible in two different ways:

First way is to impose some covariant restrictions on a super-field, to make it independent from one of two grassmannian coordinates \cite{il08}. The result is a super-field with one fermionic and one scalar field which can be used to represent a chiral supermultiplet in the theory. Such a super-field is called a "Chiral super-field" and represents scalars and matters.

Second way is to perform some generalized super-symmetric transformation on a super-field to transform away "unwanted" components \cite{bi01}. By this way, it is possible to create a super-field with a spin $1$ vector field and a Weyl spinor field. Such a super-field, which is qualified to represent a vector supermultiplet, is called "vector super-field".

\subsection{Chiral super-field}

We define a chiral super-field through following subsidiary conditions:
	 \b
	 D_i \tilde{\phi}=0
	 \; , \;\;\; \tilde{D}_i \phi=0 .
	 \e
This can be solved easily, observing that
	 $$
	 D_i \tilde{\theta} = \tilde{D}_i \theta =0
	 \; , \;\;\; \tilde{D}_i y^{\alpha} = D_i \tilde{y}^{\alpha} = 0,
	 $$
with
	 \b
	 y^{\alpha} = x^{\alpha} + i \theta \sigma^{\alpha} \tilde{\theta}
	 \; , \;\;\; \tilde{y}^{\alpha} = x^{\alpha} - i \theta \sigma^{\alpha} \tilde{\theta} .
	 \e
Hence, $\phi$ depends only on $y^{\alpha}$ and $\theta$ (i.e. all $\tilde{\theta}$ dependence is through  $y^{\alpha}$) also $\tilde{\phi}$ depends only on $\tilde{y}^{\alpha}$ and $\tilde{\theta}$.
	
Then, the chiral super-field withe condition $(4.18)$ is \cite{bi01}:
	 \b
	 \phi(y^{\alpha} , \theta) = 															 						 		 z(y^{\alpha})+\sqrt{2}\theta\psi(y^{\alpha})
	 -\theta\theta f(y^{\alpha}),
	 \e
and
	 \b
	 \tilde{\phi}(\tilde{y}^{\alpha} , \tilde{\theta}) = 															 		 \tilde{z}(\tilde{y}^{\alpha})
	 + \sqrt{2}\tilde{\theta}\tilde{\psi}(\tilde{y}^{\alpha})
	 - \tilde{\theta}\tilde{\theta} \tilde{f}(\tilde{y}^{\alpha}).
	 \e

Utilizing coordinates $y^{\alpha}$ and $\theta$, spinorial generators in this representation $(4.20)$ take the following forms:
	 \b
	 Q_i \phi \equiv \left(
	 - i \sigma^{\alpha}_{ij} y_{\alpha} {\partial \over \partial \theta_j}
	 \right) \phi ,
	 \e
	 \b
	 \tilde{Q}_i \phi \equiv \left(
	  i \sigma^{\alpha}y_{\alpha} {\partial \over \partial \tilde{\theta}_i}
	 - 2 \theta_j \sigma_{ji}^{\alpha}  \partial^T_{y^{\alpha}}
	 \right) \phi ,
	 \e

 One can rewrite the covariant derivatives as
	 \b
	 D_i =
	 \sigma^{\alpha}_{ij} y_{\alpha} {\partial \over \partial \theta_j} ,
	 \e
	 \b
	 \tilde{D}_i =
	 \sigma^{\alpha} y_{\alpha} {\partial \over \partial \tilde{\theta}_i}
	 + 2i \theta_j \sigma_{ji}^{\alpha}  \partial^T_{y^{\alpha}} .
	 \e
We also need to define a super-potential $W(\phi)$, which may be a function of several $\phi$'s:
	 \b
	 W(\phi^k) = W \left( z^k(y^{\alpha})+\sqrt{2}\theta\psi^k(y^{\alpha})
	 -\theta\theta f^k(y^{\alpha}) \right),
	 \e
where the label $k$ varies on the number of different $\phi$'s which are contributing in $W(\phi)$. Using Taylor expansion, $(4.26)$ can be written in the following form:
	 \b
	 W(\phi) = W(z) + \sqrt{2}\theta{\partial W\over \partial z}\psi
	 - \theta\theta \left( {\partial W\over \partial z}f
	 + {1\over 2}{\partial^2 W \over \partial z^k \partial z^q}\psi^k\psi^q \right) ,
	 \e
which is clearly a chiral super-field.

Any chiral super-field under infinitesimal super-symmetric transformation transforms to a chiral super-field. Applying liner extension of an infinitesimal super-symmetric transformation on $\phi(y^{\alpha},\theta)$, we have
	 $$
	 (1+i\epsilon Q+i\tilde{\epsilon}\tilde{Q}-i x^{\alpha}\sigma^{\beta}L_{\alpha\beta})
	 \phi(y^{\alpha},\theta) = \phi(y^{\alpha},\theta)
	 + \delta_{x,\epsilon,\tilde{\epsilon}}\phi .
	 $$
Using $(4.16)$, $(4.22)$ and $(4.23)$ in above equation, we obtain
	 \b
	 \delta_{x,\epsilon,\tilde{\epsilon}}\phi =
	 \left[
	 \sqrt{2} \epsilon \sigma^{\alpha} x_{\alpha} \psi
	 + x^{\alpha} \sigma^{\beta} x_{\beta} \partial^T_{\alpha}z
	 \right]
	 +\sqrt{2}\theta \left[
	 -\sqrt{2}i\sigma^{\alpha} \tilde{\epsilon} \partial^T_{\alpha}z
	 + x^{\alpha} \sigma^{\beta} x_{\beta} \partial^T_{\alpha} \psi
	 - {i \over 2} x^{\alpha} \sigma_{\alpha} \psi
	 \right]
	 $$$$
	 -\theta\theta \left[
	 2i \partial_{\alpha} \psi \sigma^{\alpha} \tilde{\epsilon}
	 + x^{\alpha} \sigma^{\beta} x_{\beta} \partial^T_{\alpha} f
	 + {i \over 2} x^{\alpha} \sigma_{\alpha} f
	 \right],
	 \e
and by comparing $(4.28)$ with $(4.20)$, the variations of the fields are:
	 $$
	 \delta z=\sqrt{2} \epsilon \sigma^{\alpha} x_{\alpha} \psi
	 + x^{\alpha} \sigma^{\beta} x_{\beta} \partial^T_{\alpha}z ,$$
	 \b
	 \delta \psi =
	 -\sqrt{2}i\sigma^{\alpha} \tilde{\epsilon} \partial^T_{\alpha}z
	 + x^{\alpha} \sigma^{\beta} x_{\beta} \partial^T_{\alpha} \psi
	 - {i \over 2} x^{\alpha} \sigma_{\alpha} \psi	,
	 \e
	 $$
	 \delta f = 2i \partial^T_{\alpha} \psi \sigma^{\alpha} \tilde{\epsilon}
	 + x^{\alpha} \sigma^{\beta} x_{\beta} \partial^T_{\alpha} f
	 + {i \over 2} x^{\alpha} \sigma_{\alpha} f .
	 $$

\subsection{Vector super-field}	
	
A vector super-field can be define in representation $(4.7)$ as \cite{bi01}:
	 \b
	 V(x^{\beta} ,\theta,\tilde{\theta}) =
	 A(x^{\beta}) + \theta\chi(x^{\beta})
	 + \tilde{\theta}\tilde{\chi}(x^{\beta})
	 + \theta\theta \left[ m(x^{\beta})+in(x^{\beta})\right]
	 + \tilde{\theta}\tilde{\theta} \left[ m(x^{\beta})-in(x^{\beta})\right]
	 $$$$
	 + \theta\sigma^{\alpha}\tilde{\theta}\nu_{\alpha}(x^{\beta})
	 + \theta\theta\tilde{\theta}\tilde{\lambda}(x^{\beta})
	 + \tilde{\theta}\tilde{\theta}\theta\lambda(x^{\beta})
	 + \theta\theta\tilde{\theta}\tilde{\theta}B(x^{\beta}) .
	 \e	
There are 8 bosonic components ($A,B,m,n, \nu_{\alpha}$ with the condition $x.\nu=0$) and 8 fermionic components ($\chi$, $\lambda$ and their complex conjugations). These are too many components
to describe a single supermultiplet. We want to reduce their number
by making use of the super-symmetric generalisation of a gauge transformation.
Note that the transformation \cite{bi01}	
	 \b
	 V \longrightarrow V-(\Lambda + \tilde{\Lambda}),
	 \e
with $\Lambda$ as a chiral super-field is an abelian gauge transformation. We conclude that $(4.31)$ is our desired
super-symmetric generalisation. If this transformation, $(4.31)$, is a symmetry (actually a gauge symmetry) of the theory, then, by an appropriate choice of $\Lambda$, one can transform away the components $\chi,A,m,n$ and one component of $\nu_{\alpha}$. This choice is called the Wess-Zumino gauge, and it reduces the
vector super-field to
	 \b
	 V_{WZ}
	 = \theta\sigma^{\alpha}\tilde{\theta}\nu_{\alpha}(x^{\beta})
	 + \theta\theta\tilde{\theta}\tilde{\lambda}(x^{\beta})
	 + \tilde{\theta}\tilde{\theta}\theta\lambda(x^{\beta})
	 + \theta\theta\tilde{\theta}\tilde{\theta}B(x^{\beta}) .
	 \e

To construct kinetic terms for the vector field $\nu_{\alpha}$, we must act on $V_{WZ}$, with covariant derivatives  $D_i$ and $\tilde{D}_i$, $(4.16)$ and $(4.17)$, through definition
	 \b
	 W_i=-{1 \over 4}\tilde{D}\tilde{D}(e^{-V_{WZ}}D_ie^{V_{WZ}}).
	 \e
Considering $(D)^3=(\tilde{D})^3=0$, it is obvious that
	 $$
	 \tilde{D}_i W^i=0.
	 $$
$W_i$ is clearly a chiral super-field, which transforms under $(4.31)$, as
	 \b
	 W_i \longrightarrow e^{\Lambda}W_ie^{-\Lambda} .
	 \e
According to $(4.32)$, since each component of $V_{WZ}$ contains at least one $\theta$, the only non-vanishing power of $V_{WZ}$ is
	 \b
	 (V_{WZ})^2 = {1 \over 2}\theta\theta\tilde{\theta}\tilde{\theta}\nu_{\alpha}\nu^{\alpha}
	 \e
and $(V_{WZ})^n =0$ for $n>2$. Using this property, we obtain:
	 \b
	 e^{V_{WZ}}=1 + V_{WZ}+ {1 \over 2}(V_{WZ})^2.
	 \e
By replacing $(4.36)$ in definition $(4.33)$, the following relation is obtained:
	 \b
	 W_i=-{1 \over 4}\tilde{D}\tilde{D}(D_iV_{WZ})
	 + {1 \over 8}\tilde{D}\tilde{D} \left[ D_iV_{WZ},V_{WZ} \right] .
	 \e
Hereafter, it is more convenient to use notation $V$ instead of $V_{WZ}$ to represent a vector super-field.

\setcounter{equation}{0}
\section{N=1 super-symmetry Lagrangian in de Sitter space}

A simple super-symmetry Lagrangian in the de Sitter space contains two general parts: pure gauge Lagrangian and matter Lagrangian. Pure gauge Lagrangian only involves the vector supermultiplet and can be described in term of the vector super-fields. Matter Lagrangian describes couplings among chiral matter fields with scalars and vector fields. We start with building the pure gauge Lagrangian and afterwards we present the matter Lagrangian, then, summarising these two parts, super-symmetry $N=1$ Lagrangian in the de Sitter space-time can be obtained.

As mentioned, the kinetic terms for a vector super-field could be provided through definition $(4.33)$. Using $(4.34)$, it is easy to see that $\int{d^2\theta\;W_iW^i}$ is a super-symmetry gauge invariant Lagrangian. To obtain its component expansion, we need the $\theta\theta$-terms in $W_iW^i$. Using $(4.16)$, $(4.17)$, $(4.32)$ and 												 \b
	 \tilde{D}\tilde{D}\tilde{\theta}\tilde{\theta}
	 = -4 (\tilde{D} \tilde{\theta})^2
	 = -4 (\sigma^{\alpha} x_{\alpha})(\sigma^{\beta} x_{\beta})
	 = -4x^2,
	 \e
one can obtain terms of $(4.37)$ as follows:
	 \b
	 -{1 \over 4}\tilde{D}\tilde{D}(D_iV) =
	 x^2 (\sigma^{\alpha} x_{\alpha} \lambda)_i
	 + 2 x^2 (\theta \sigma^{\alpha} x_{\alpha})_i B
	 + {i \over 4} x^2 (\theta \sigma^{\alpha} \sigma^{\beta})_i
	 (\partial_{\alpha}^T \nu_{\beta} - \partial_{\beta}^T \nu_{\alpha})
	 $$$$
	 - {i \over 2} \theta\theta x^2 (\sigma^{\alpha} \partial_{\alpha}^T \tilde{\lambda})_i,
	 \e
and
	 \b
	 {1 \over 8}\tilde{D}\tilde{D}[D_iV,V] =
	 {1 \over 4} \theta_j x^2 x^{\alpha}\sigma^{\beta}_{ij} [\nu_{\alpha},\nu_{\beta}]
	 + {1 \over 2} \theta\theta x^2 x^{\alpha} [\nu_{\alpha},\tilde{\lambda}_i]
	 $$$$
	 \hspace{2.3cm} =
	 {1 \over 4} \theta_j x^2 x^{\alpha}\sigma^{\beta}_{ij} [\nu_{\alpha},\nu_{\beta}]
	 + {1 \over 2} \theta\theta x^2 x^{\alpha} \tilde{\lambda}_i \nu_{\alpha} ,
	 \e
$V$ is defined in adjoint representation of the gauge group, so
	 $$
	V= V^a T_a ,
	 $$
	 $$
	 V^a
	 = \theta\sigma^{\alpha}\tilde{\theta}\nu_{\alpha}^a(x^{\beta})
	 + \theta\theta\tilde{\theta}\tilde{\lambda}^a(x^{\beta})
	 + \tilde{\theta}\tilde{\theta}\theta\lambda^a(x^{\beta})
	 + \theta\theta\tilde{\theta}\tilde{\theta}B^a(x^{\beta})
	 ,
	 $$
where $T^a$'s are generators of the adjoint representation of the gauge group. To simplify the method, we may adopt a gauge group with only one generator (i.e. $U(1)$). Of course, it is expansion to a larger gauge group is trivial. Now, embedding $(5.2)$ and $(5.3)$ in $(4.37)$ gives:
	 \b
	 W_i =
	 x^2 (\sigma^{\alpha} x_{\alpha} \lambda)_i
	 + 2 x^2 (\theta \sigma^{\alpha} x_{\alpha})_i B
	 + {i \over 4} x^2 \theta_i \sigma^{\alpha} \sigma^{\beta} F_{\alpha\beta}
	 - {i \over 2} \theta\theta x^2 (\sigma^{\alpha} D^{(\lambda)}_{\alpha} \tilde{\lambda})_i	
	 ,
	 \e
where
	 \b
	 F_{\alpha\beta} =
	 \partial_{\alpha}^T \nu_{\beta}
	 - \partial_{\beta}^T \nu_{\alpha}
	 - i \sigma_{\gamma} x^{\gamma} [\nu_{\alpha},\nu_{\beta}] ,
	 \e
and
	 \b
	 D_{\alpha}^{(\lambda)} \lambda \equiv
	 \partial_{\alpha}^T \lambda
	 - i \sigma^{\gamma} x_{\gamma} \lambda \nu_{\alpha} ,
	 \e
which is gauge covariant derivative for spinor field $\lambda$. Consequently, we can derive
	 \b
	 W_iW^i\vert_{\theta\theta} \; : \;
 	 - {1 \over 4} \theta\theta x^4 F_{\alpha\beta} F^{\alpha\beta}	
	 - i \theta\theta x^4 \lambda x^{\alpha} D_{\alpha}^{(\lambda)}\tilde{\lambda}
	 - 2 \theta\theta x^6 B^2
	 ,
	 \e
where $x^4= \left( x^2 \right)^2 = \left( H^{-2} \right)^2$. Then, the pure gauge Lagrangian is
	 \b
	 L_{gauge}=
 	 - {1 \over 4} x^4 F_{\alpha\beta} F^{\alpha\beta}	
	 - i x^4 \lambda x^{\alpha} D_{\alpha}^{(\lambda)}\tilde{\lambda}
	 - 2 x^6 B^2 .
	 \e

A matter field will transform under gauge transformation by the following form:
	 \b
	 \phi^i \longrightarrow (e^\Lambda)^i_j \phi^j \; , \;\;\;
	 \tilde{\phi}^i \longrightarrow \tilde{\phi}^j (e^{\tilde{\Lambda}})^i_j .
	 \e
Considering this transformation law and $(4.31)$, the term $\tilde{\phi} e^V \phi$ is invariant under gauge transformation:
	 \b
	 \tilde{\phi} e^V \phi \longrightarrow \tilde{\phi} e^{\tilde{\Lambda}}
	 (e^{-\tilde{\Lambda}} e^V e^{-\Lambda}) e^{\Lambda} \phi =	
	 \tilde{\phi} e^V \phi .
	 \e
So, a simple form for the super-symmetry invariant Lagrangian for matter fields may be
	 \b
	 L_{matter} = \int d^2\theta \; d^2\tilde{\theta} \; \tilde{\phi} e^V \phi
	 + \int d^2\theta \; W(\phi) + \int d^2\tilde{\theta} \; \tilde{W}(\tilde{\phi}) ,
	 \e
where $W(\phi)$ and $\tilde{W}(\tilde{\phi})$ are chiral super-fields $[2]$. Note that any Lagrangian of the form
	 $$
	 \int d^2\theta \; d^2\tilde{\theta} \; F(x^{\alpha}, \theta, \tilde{\theta})
	 + \int d^2\theta \; W(\phi) + \int d^2\tilde{\theta} \; \tilde{W}(\tilde{\phi})
	 $$
is automatically super-symmetry invariant (i.e. it transforms at most by a total derivative in space-time), where $F$ is a super-field by definition $(4.10)$. Now, we need to compute the $\theta \theta \tilde{\theta} \tilde{\theta}$ components of $\tilde{\phi} e^V \phi$:
	 \b
	 \tilde{\phi} e^V \phi = \tilde{\phi} \phi + \tilde{\phi} V \phi + {1 \over 2} \tilde{\phi} V^2 \phi .
	 \e
In case of mixing the chiral super-fields $\phi$ and $\tilde{\phi}$ with the vector super-field $V$ on $(5.12)$, first it is necessary to bring $\phi$ and $\tilde{\phi}$ into representation $(4.7)$. Using $(4.9)$, it is easy to show that
	 \b
	 \phi( x^{\beta} ,\theta,\tilde{\theta}) =
	 z(x^{\beta})
	 + i \theta \sigma^{\alpha} \tilde{\theta} \partial_{\alpha}^T z(x^{\beta})
	 - {1 \over 4} \theta\theta \tilde{\theta}\tilde{\theta} (\partial^T)^2 z(x^{\beta})
	 + \sqrt{2}\theta\psi(x^{\beta})
	 $$$$
	 - {i \over \sqrt{2}} \theta\theta \partial_{\alpha}^T \psi \sigma^{\alpha} \tilde{\theta}
	 - \theta\theta f(x^{\beta}) ,
	 \e
	 \b
	 \tilde{\phi}( x^{\beta} ,\theta,\tilde{\theta}) =
	 \tilde{z}(x^{\beta})
	 - i \theta \sigma^{\alpha} \tilde{\theta} \partial_{\alpha}^T \tilde{z}(x^{\beta})
	 - {1 \over 4} \theta\theta \tilde{\theta}\tilde{\theta} (\partial^T)^2 \tilde{z}(x^{\beta})
	 + \sqrt{2} \tilde{\theta} \tilde{\psi}(x^{\beta})
	 $$$$
	 + {i \over \sqrt{2}} \tilde{\theta} \tilde{\theta} \theta \sigma^{\alpha} \partial_{\alpha}^T \tilde{\psi}
	 - \tilde{\theta} \tilde{\theta} \tilde{f}(x^{\beta}) .
	 \e
From $(5.13)$, $(5.14)$ and $(4.32)$, for $(5.12)$ we obtain:
	 \b
	 \tilde{\phi} \phi \vert_{\theta\theta\tilde{\theta}\tilde{\theta}}
	 = \partial_{\alpha}^T \tilde{z} \partial^{T \alpha} z
	 - i \psi \sigma^{\alpha} \partial_{\alpha}^T \tilde{\psi}
	 + \tilde{f} f
	 + \mbox{total derivatives} ,
	 \e
	 \b
	 \tilde{\phi} V \phi \vert_{\theta\theta\tilde{\theta}\tilde{\theta}}
	 = {i \over 2} \tilde{z} \nu^{\alpha} \partial_{\alpha}^T z
	 - {i \over 2} \left( \partial_{\alpha}^T \tilde{z} \right) \nu^{\alpha} z
	 - {\sqrt{2} \over 2} \tilde{z} \lambda \psi
	 - {\sqrt{2} \over 2} \tilde{\psi} \tilde{\lambda} z
	 - \tilde{\psi} \sigma^{\alpha} \nu_{\alpha} \psi
	 + \tilde{z} B z
	 ,
	 \e
and
	 \b
	 \tilde{\phi} V^2 \phi \vert_{\theta\theta\tilde{\theta}\tilde{\theta}}
	 = {1 \over 2} \tilde{z}\nu^{\alpha}\nu_{\alpha}z .
	 \e
The terms which are marked as "total derivatives" in $(5.15)$ will be vanished during integration over all space-time. Also from $(4.27)$, we can conclude
	 \b
	 W(\phi)\vert_{\theta\theta} =
	 - {\partial W\over \partial z}f
	 - {1\over 2}{\partial^2 W \over \partial z^k \partial z^q} \psi^k \psi^q,
	 \e
and
	 \b
	 \tilde{W}(\tilde{\phi})\vert_{\tilde{\theta}\tilde{\theta}} =
	 - {\partial \tilde{W}\over \partial \tilde{z}}\tilde{f}
	 - {1\over 2}{\partial^2 \tilde{W} \over \partial \tilde{z}^k \partial \tilde{z}^q}\tilde{\psi}^k\tilde{\psi}^q
	 .
	 \e
Now, we can put above relations to gather and obtain the matter Lagrangian $(5.11)$:
	 \b
	 L_{matter}
	 = (\tilde{D}^{(z)}_{\alpha} \tilde{z})(D^{(z)\alpha}z)
	 - i \psi \sigma^{\alpha} D^{(\psi)}_{\alpha} \tilde{\psi}
	 - {\sqrt{2} \over 2} \tilde{z} \lambda \psi
	 - {\sqrt{2} \over 2} \tilde{\psi} \tilde{\lambda} z	
	 + \tilde{f} f
	 - {\partial W\over \partial z}f	
	 - {\partial \tilde{W}\over \partial \tilde{z}}\tilde{f}
	 $$$$
	 + \tilde{z} B z	
	 - {1\over 2}{\partial^2 W \over \partial z^k \partial z^q}\psi^k\psi^q
	 - {1\over 2}{\partial^2 \tilde{W} \over \partial \tilde{z}^k \partial \tilde{z}^q}\tilde{\psi}^k\tilde{\psi}^q
	 + \mbox{total derivatives}
	 ,
	 \e
where
	 \b
	 D^{(z)}_{\alpha} z \equiv
	 \partial_{\alpha}^T z
	 - {i \over 2} \nu_{\alpha} z ,
	 \e
	 \b
	 D^{(\psi)}_{\alpha} \psi \equiv
	 \partial_{\alpha}^T \psi
	 + i \nu_{\alpha} \psi ,
	 \e
are gauge covariant derivatives for respectively scalar field $z$ and spinor field $\psi$. The reader should not confuse the gauge covariant derivatives, $(5.6)$, $(5.21)$, and $(5.22)$, with the super-covariant derivatives $D_i$ and $\tilde{D}_i$.

Finally, the super-symmetry invariant Lagrangian in the de Sitter space-time can be defined as:
	 \b
	 L_{N=1}^{ds} = L_{gauge} + L_{matter}.
	 \e
Using $(5.8)$ and $(5.20)$, $(5.23)$ can be written as follows:
	 \b
	 L_{N=1}^{ds}
	 = - {1 \over 4} x^4 F_{\alpha\beta} F^{\alpha\beta}	
	 - i x^4 \lambda x^{\alpha} D_{\alpha}^{(\lambda)}\tilde{\lambda}
	 + (\tilde{D}^{(z)}_{\alpha} \tilde{z})(D^{(z)\alpha}z)
	 - i \psi \sigma^{\alpha} D^{(\psi)}_{\alpha} \tilde{\psi}
	 - {\sqrt{2} \over 2} \tilde{z} \lambda \psi
	 - {\sqrt{2} \over 2} \tilde{\psi} \tilde{\lambda} z
	 $$$$	
	 - {1\over 2}{\partial^2 W \over \partial z^k \partial z^q}\psi^k\psi^q
	 - {1\over 2}{\partial^2 \tilde{W} \over \partial \tilde{z}^k \partial \tilde{z}^q}\tilde{\psi}^k\tilde{\psi}^q
	 - R(z,\tilde{z}) +\mbox{total derivatives}
	 ,
	 \e
where
	 \b
	 R(z, \tilde{z})
	 = 2 x^6 B^2
	 - \tilde{z} B z	
	 - \tilde{f} f
	 + {\partial W\over \partial z}f	
	 + {\partial \tilde{W}\over \partial \tilde{z}}\tilde{f}
	 \e
is the scalar potential. In absence of any kinetic terms for scalar fields $f$ and $B$ in $(5.25)$, one can conclude that these are auxiliary fields and we need to remove them by making some proper choices. These choices can be
	 \b
	 f = {\partial \tilde{W} \over \partial \tilde{z}} \; ,\;\;\;
	 B = {\tilde{z}z \over x^6 } .
	 \e
Then, the scalar potential $R(z, \tilde{z})$ takes the form
	 \b
	 R(z, \tilde{z})
	 = x^6 B^2	
	 +  \tilde{f} f	.
	 \e
These results are highly alike to the Minkowskian Lagrangian for $N=1$ super-symmetry $[2]$. One can easily extract the equations of motion of the different fields, which are appeared in the Lagrangian $(5.24)$ and compare them with the equations of motion of the original fields in de Sitter space-time.

\section{ Conclusions}

In a series of papers, it has been shown that the formalism of the quantum field in the de Sitter universe, in ambient space notation, is very similar to the quantum field formalism in Minkowski space-time. In this frame, it is proven that, super-algebra with odd $N$ can be closed in the de Sitter ambient space notation \cite{PRT05} similar to the Minkowskian space. This paper, successfully introduces the super-symmetry Lagrangian for $N=1$ in de Sitter ambient space notation, $(5.24)$. This formalism is highly alike to the Minkowskian counterpart and a very simple relative to intrinsic coordinates system. The importance of this formalism may be shown further by the consideration of the linear conformal gravity \cite{tapeta12} and super-conformal gravity in the de Sitter space-time, which have the important role for considering the quantum gravity and its unification with the other interactions.

\end{document}